\documentclass[a4paper,11pt]{article}
\usepackage{pos}

\usepackage{graphicx} 
 
\title{
CASK: A Gauge Covariant Transformer for Lattice Gauge Theory
}

\author[a]{Yuki Nagai}
\author[b]{Hiroshi Ohno}
\author*[c,d]{Akio Tomiya}

\affiliation[a]{Information Technology Center, University of Tokyo, Kashiwa, Chiba 277–0882, Japan}
\affiliation[b]{Department of Advanced Materials Science, University of Tokyo, Kashiwa, Chiba 277-8561, Japan}

\affiliation[c]{Center for Computational Sciences, University of Tsukuba, Tsukuba, Ibaraki 305-8577, Japan}
\affiliation[d]{Department of Mathematics, Tokyo Woman’s Christian University, Tokyo 167-8585, Japan}
\affiliation[e]{RIKEN Center for Computational Science, Kobe 650-0047, Japan}

\emailAdd{nagai.yuki@mail.u-tokyo.ac.jp}
\emailAdd{hohno@ccs.tsukuba.ac.jp}
\emailAdd{akio@yukawa.kyoto-u.ac.jp}

\abstract{
We propose a Transformer neural network architecture specifically designed for lattice QCD, focusing on preserving the fundamental symmetries required in lattice gauge theory. The proposed architecture is gauge covariant/equivariant, ensuring it respects gauge symmetry on the lattice, and is also equivariant under spacetime symmetries such as rotations and translations on the lattice.
A key feature of our approach lies in the attention matrix, which forms the core of the Transformer architecture. To preserve symmetries, we define the attention matrix using a Frobenius inner product between link variables and extended staples. This construction ensures that the attention matrix remains invariant under gauge transformations, thereby making the entire Transformer architecture covariant.
We evaluated the performance of the gauge covariant Transformer in the context of self-learning HMC. Numerical experiments show that the proposed architecture achieves higher performance compared to the gauge covariant neural networks, demonstrating its potential to improve lattice QCD calculations.
}

\FullConference{The 41st International Symposium on Lattice Field Theory (LATTICE2024)\\
 28 July - 3 August 2024\\
Liverpool, UK\\}

\begin{document}
\maketitle

\section{Introduction}
Machine learning has become a powerful tool in high energy physics, where the computational cost of large-scale simulations is often prohibitively high. One promising approach involves the use of neural surrogate models, which serve as cheaper approximations to the exact theory and can significantly reduce computational cost \cite{Adelmann:2022ozp}. 
In particular, gauge covariant/equivariant neural networks\footnote{
The conceptual foundations of these architectures draw on the distinction between equivariance and covariance \cite{Marcos_2017}.
} are attracting considerable attention because they allow flexible and differentiable mappings between gauge fields, controlled by learnable parameters. 

Machine learning (ML) techniques have recently made significant inroads into lattice QCD, offering novel strategies to tackle long-standing computational challenges. One notable direction is the development of flow-based sampling algorithms \cite{Albergo:2019eim,Kanwar:2020xzo,Boyda:2020hsi,Albergo:2021bna,Albergo:2022qfi,Abbott:2022zhs,Abbott:2024kfc} and continuous flow approaches \cite{deHaan:2021erb,Gerdes:2022eve,Gerdes:2024rjk}, which promise more efficient generation of gauge field configurations. ML has also begun to play a key role in non-equilibrium Monte Carlo \cite{Bulgarelli:2024yrz,Bulgarelli:2024cqc,Caselle:2024ent}. Furthermore, perfect action techniques are being refined through advanced neural network methods \cite{Holland:2024muu}, and preconditioning strategies employing neural networks have been shown to enhance numerical efficiency \cite{Lehner:2023bba,Lehner:2023prf,Cali:2022qbd}. Taken together, these innovations highlight the rapidly evolving synergy between ML and lattice QCD, potentially broadening the scope of feasible calculations.

In order for a neural network architecture to be useful in lattice QCD, it must satisfy several important criteria without sacrificing either efficiency or physical rigor. It must be compatible with gauge symmetry, as gauge invariance is a cornerstone of QCD and its lattice formulation \cite{cuomo2022scientificmachinelearningphysicsinformed}. It must also be “fermion friendly” and accommodate modern lattice QCD simulations that incorporate dynamical fermions. Furthermore, the architecture must be fully differentiable to allow training via gradient-based methods, which have shown immense utility in machine learning \cite{ruder2017overviewgradientdescentoptimization,blondel2024elementsdifferentiableprogramming,moses2020instead}.

Recent breakthroughs in deep learning suggest that Transformers, originally popularized in natural language processing \cite{vaswani2023attentionneed} and known for their ability to capture non-local correlations, can be beneficial for problems in lattice QCD. Their core technology, the attention matrix, can handle long-range interactions \cite{lin2021surveytransformers}, a feature particularly relevant in the presence of fermions. There is also growing interest in exploiting symmetries within the data, motivating research on how to scale equivariant architectures with Transformers \cite{brehmer2024doesequivariancematterscale}.

Despite these promising directions, constructing a network that is both gauge covariant and capable of leveraging the flexibility of the Transformer paradigm remains a challenge. In this work, we address these issues by introducing a new neural network layer, \emph{\underline{C}ovariant \underline{A}ttention with \underline{S}tout \underline{K}ernel} (CASK), which serves as a gauge covariant attention block. Our approach integrates the requirements of gauge covariance, differentiability, and fermion friendliness, while harnessing the ability of Transformer architectures to learn and exploit non-local correlations in lattice QCD. 

\section{Gauge covariant Transformer (CASK)}
We first provide an overview of the gauge covariant neural network formalism, 
which serves as the foundation for constructing CASK.

\subsection{Gauge covariant neural network}
We first recall the gauge covariant neural network \cite{Nagai:2021bhh}, which can be viewed as a trainable version of the stout smearing commonly used in lattice gauge theory. The gauge covariant neural network is also known as a residual flow \cite{Abbott:2024kfc}.
In the following discussion, we focus on the simplest version of this gauge covariant neural network, namely the stout-type construction, which can be regarded as a convolutional layer acting on the links of the lattice gauge field.

To process gauge fields in a covariant manner, we consider an iterative evolution equation for the link variables $U_\mu^{(l)}(n) \in \mathrm{SU}(N_c)$ of the form
\begin{align}
U_\mu^{(l+1)}(n)=
g^{(l)}_{n,\mu}
U_\mu^{(l)}(n),
\end{align}
where the gauge update
\begin{align}
g^{(l)}_{n,\mu}=\exp\Bigl[{\rm i} \sum_{f} \rho^{(l,f)} Q_\mu^{(l,f)}(n)\Bigr] \,\in\, \mathrm{SU}(N_c)
\end{align}
encodes the action of the neural network at layer $l$. 
Here, 
$l$ is the number of layers (or smearing levels), and $f$ indicates the type of loops considered (for example, staples of various shapes). The real parameters $\rho^{(l,f)} \in \mathbb{R}$ are trainable weights. 

Each $Q^{(l,f)}_\mu(n)$ is an element of the Lie algebra $\mathfrak{su}(N_c)$ constructed from a closed loop $\Omega_\mu(n)$ surrounding the link $U_\mu(n)$. Concretely,
\begin{align}
Q_\mu(n)=\frac{\mathrm{i}}{2}\Bigl(\Omega_\mu^{\dagger}(n)-\Omega_\mu(n)\Bigr)\;-\;\frac{\mathrm{i}}{2\,N_c}\,\mathrm{Tr}\Bigl(\Omega_\mu^{\dagger}(n)-\Omega_\mu(n)\Bigr)
\in
\mathfrak{su}(N_c)
,
\label{eq:AT-loop}
\end{align}
where $\Omega_\mu(n)\in {\rm SU}(N_c)$ is formed by the product of links associated with $U_\mu(n)$. In essence, this construction implements a local transformation that “smears” or modifies the original link $U_\mu^{(l)}(n)$ to produce $U_\mu^{(l+1)}(n)$, while ensuring that the operation remains gauge covariant. By design, this network preserves gauge symmetry and can be trained using a generalized backpropagation scheme adapted for matrix-valued variables.

Viewed this way, the stout-type gauge covariant neural network recasts a well-known smearing procedure as a trainable, parameterized transformation. 
This perspective not only bridges lattice smearing methods and modern deep learning, 
but also connects them in a unified framework.

\subsection{Lesson from Transformer for spins}
Before introducing gauge covariant Transformer,
here we briefly review transformer for a classical $\mathrm{O}(3)$ spin model with quantum electrons in two dimensions \cite{Nagai:2023fxt,Tomiya:2023jdy}. 
This is helpful to understand gauge covariant Transformer.
Let $\vec{S}_{n} \in \mathbb{R}^3$ be a scalar field on lattice, which is a component of a classical spin. 
Here $n$ indicates lattice site.
Spin variables are normalized as $\sum_{\mu=1}^3 |\vec{S}_{n}|^2  = 1$ for all $n$.
The Hamiltonian of the system is invariant under a spin rotation $\vec{S}_{n} \to R \vec{S}_{n}$ with $R \in \mathrm{O}(3)$.
This transformation is independent of coordinate $n$ (global symmetry).

As a first step of the procedure, we perform three different block spin transformations with different weights. The transformation is,
\begin{align}
\vec{S}^{(\alpha)}_{n}
=
\sum_{n' \in N^{(k)}_n} w^{(\alpha)}_k
\vec{S}_{n'},
\end{align}
for $\alpha = {\rm Q,K,V}$
and $w^{(\alpha)} \in \mathbb{R}$ is a weight.
$N_n^{(k)}$ indicates a set of $k$-th neighbors for a lattice site $n$.
We remark that this is covariant (equivariant) under global $\mathrm{O}(3)$ transformation for spin $S_{\mu,n}$.
Next we construct an attention matrix.
By using the standard inner product for two real vectors,
\begin{align}
\tilde{M}_{n'n} 
= \sum_{\mu}
\vec{S}^{\rm (K)}_{n'}
\cdot
\vec{S}^{\rm (Q)}_{n}
= \sum_{\mu}
\big(\vec{S}^{\rm (K)}_{n'} \big)^\top
\vec{S}^{\rm (Q)}_{n},\;\;\;
M_{n'n} = {\rm ReLu}(\tilde{M}_{n'n})
\end{align}
Here $M_{n'n} \in \mathbb{R}_{+}$ is called an {\rm attention matrix}
and ${\rm ReLu}$ is the rectified linear function\footnote{
In the original Transformer uses softmax function instead of ReLu.
ReLu helps to keep symmetry and numerically cheaper than the softmax.
The latter is philosophically similar to the flash attention.
}.
$M_{nn'}$ is obviously related to correlation functions.
This matrix connects correlations between all points to all points, so it is a dense matrix. 
We emphasize that this is {\it invariant} under global $\mathrm{O}(3)$ transformation for spin $\vec{S}_{n}$.
Next we construct a spin operator with attention,
\begin{align}
\vec{S}^{\rm (A)}_{n} = \sum_{n'} M_{nn'}\vec{S}^{\rm (V)}_{n'},
\end{align}
and this is {\it covariant/equivariant} under global $\mathrm{O}(3)$ transformation. 
This contains correlations from all points from this system.
Finally, we construct output of the self-attention block,
\begin{align}
\vec{S}_{n}' = 
\mathcal{N}(\vec{S}_{n} + \vec{S}^{\rm (A)}_{n} )
\end{align}
$\mathcal{N}(\cdot)$ is a normalization operation to keep length of output vector to be one point-wisely. 
In neural network language, this is a layer normalization.
This is equivariant under global $\mathrm{O}(3)$ as well.
Whole procedure can be nested, which makes the network deeper.

Lessons are as follows.
The attention matrix is essentially a correlation function, which allows us to capture long-range correlations.
The attention matrix should be invariant under the symmetry transformation. So the output of the attention operation is covariant.
Output should be normalized, otherwise we cannot regard the output of a Transformer block as a spin configuration.
The normalization operation helps the training because we initialize weights with nearly zero and the attention block behaves as an identity. This enables greedy layer-wise training \cite{belilovsky2019greedylayerwiselearningscale}.

\subsection{CASK (Gauge covariant Transformer)}
Here we introduce CASK, which is a new Transformer specifically designed to process gauge configurations in a manner analogous to the covariant attention for spin systems, but with additional structure so that CASK is covariant under local gauge transformations. CASK can be regarded as a synthesis of two core ideas, namely the gauge covariant neural network and the $\mathrm{O}(3)$ equivariant Transformer. The central challenge in constructing a gauge covariant Transformer is the design of an invariant attention matrix.

\begin{figure}[t]
\begin{center}
\includegraphics[scale=0.43]{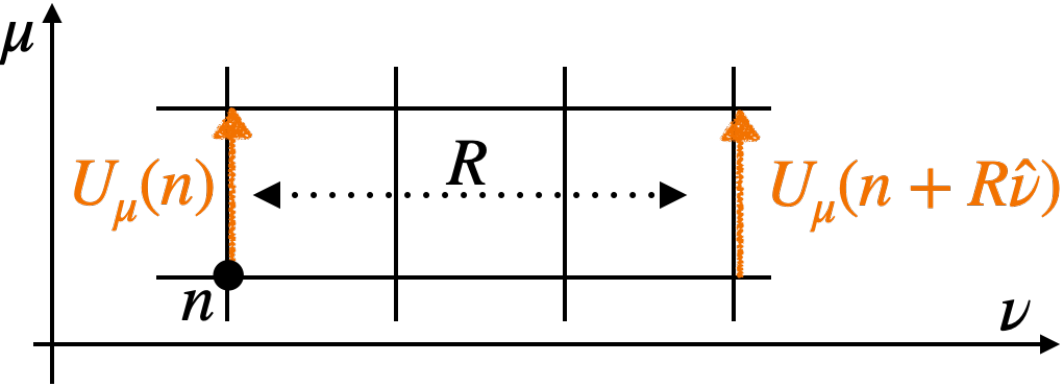}
\hspace{5mm}
\includegraphics[scale=0.43]{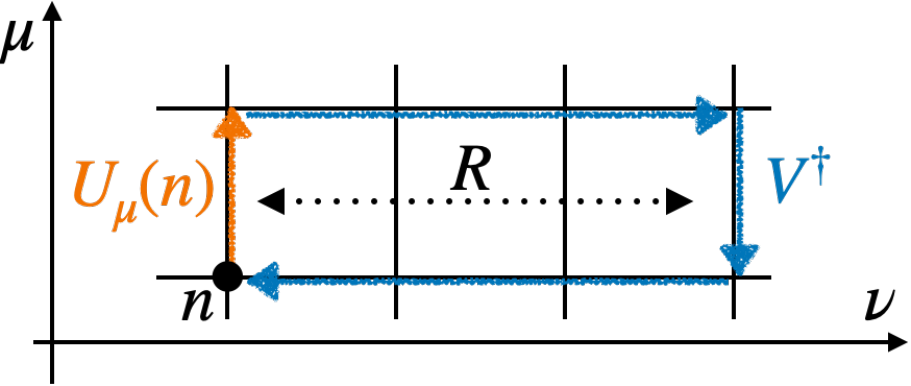}
\end{center}
\caption{Construction of the attention matrix. {\it (Left):} Two links separated by $R$. {\it (Right):} Gauge covariant combination of two links separated by $R$. \label{fig:Correlation for two links}}
\end{figure}

To achieve an invariant attention matrix, we employ the Frobenius inner product $\mathrm{tr}(A^\dagger B)$ for complex matrices $A$ and $B$. This quantity is invariant under the transformation $A \to \omega_1 A \omega_2$ and $B \to \omega_1 B \omega_2$ for $\omega_1,\omega_2 \in {\rm SU}(N_c)$, because
\[
\mathrm{tr}(A^\dagger B) \;\to\; \mathrm{tr}\bigl(\omega_2^\dagger A^\dagger \omega_1^\dagger \,\omega_1 B \omega_2\bigr)\;=\;\mathrm{tr}(A^\dagger B).
\]
The attention matrix in a Transformer is a collection of pairwise correlations, we would like to construct the attention matrix from gauge symmetric two-point functions. A naive extension from the spin case might consider the two-point correlation of two links, as shown in the left panel of Fig.~\ref{fig:Correlation for two links}. However, that combination is not gauge symmetric. Instead, one must employ a Wilson loop, illustrated in the right panel of Fig.~\ref{fig:Correlation for two links}, which ensures gauge symmetry.

The definition of the CASK layer is as follows.
To define the Transformer, we need to define three kinds of vectors.
Smeared gauge links 
$U^{(Q)}$, $U^{(K)}$ and $U^{(V)}$, corresponding to the ``query'', ``key'' and ``value'' vectors, are defined as 
\begin{align}
&
U^{(Q)}_{\mu}(n) \equiv U^{(\alpha)}_{\mu}(n;  \rho^{(\alpha)}  )\Bigg|_{\alpha\to Q},\;
U^{(K)}_{\mu}(n) \equiv U^{(\alpha)}_{\mu}(n;  \rho^{(\alpha)}  )\Bigg|_{\alpha\to K},
\;
U^{(V)}_{\mu}(n) \equiv U^{(\alpha)}_{\mu}(n;  \rho^{(\alpha)}  )\Bigg|_{\alpha\to V}.
\end{align}
These are defined by gauge covariant layer using single plaquette with independent weights $\rho^{(\alpha)}$ in this work.

For general links field $U$,
the extended staples $V_{\nu, n+\hat\mu;s}( \{ U \} )$ as the functional of $\{ U \}$ are defined as 
\begin{align}
V_{\nu, n+\hat\mu;s}( \{ U \} ) 
&\equiv
\Bigg(\prod_{t=0}^{s-1} U_{\nu}( n+\hat\mu + t\hat\nu) \Bigg)
U_{\mu}^{\dagger}(n+\hat\mu+s\hat\nu)
\Bigg(\prod_{t=0}^{s-1} U_{\nu}^{\dagger}( n+ (s-1-t)\hat\nu) \Bigg).
\end{align}
By using the staples, the attention matrix $a_{n,\mu,\nu,s}$ is defined as 
\begin{align}
a_{n,\mu,\nu,s} &= \tan \left( \frac{4}{N_c}  \tilde{a}_{n,\mu,\nu,s}  \right), \\
\tilde{a}_{n,\mu,\nu,s} &= {\rm Re} \: {\rm Tr} \left[ U^{(Q)}_{\mu}(n) V_{\nu, n+\hat\mu;s} ( \{ U^{(K)} \}) \right] - {\rm Re} \: {\rm Tr} \left[ U_{\mu}(n) V_{\nu, n+\hat\mu;s} ( \{ U \}) \right] \label{eq:attention-tilde},
\end{align}
where the second term of \eqref{eq:attention-tilde} is the input of the $l$-th CASK layer.
To amplify the signal, we introduce the tangent function.
In principle, one could construct an all-to-all attention matrix, but this work uses a sparse attention matrix \cite{child2019generatinglongsequencessparse,lin2021surveytransformers} to reduce numerical cost. Specifically, links are connected within $1\times1$, $1\times2$ and $1\times3$ rectangular Wilson loops\footnote{
This part is related to the definition of $R$ in later sentence.
}. 
This attention matrix is gauge invariant.

A relation between $l+1$-th and $l$-th CASK layer is expressed as 
\begin{align}
U^{(l+1)}_{\mu}(n) &\equiv e^{{\rm i} Q_{\mu}^{A}(n; \{ \rho \delta_{\mu\nu} \} )} U_{\mu}^{(l)}(n) 
\end{align}
where
\begin{align}
     Q_{\mu}^{A}(n; \{  a_{n,\mu,\nu,s} \} ) &= \frac{{\rm i}}{2}(\Omega_{\mu}^{A}(n; \{  a_{n,\mu,\nu,s}  \} )- \Omega^{A \dagger}_{\mu}(n; \{  a_{n,\mu,\nu,s} \} )) \nonumber \\
     &- \frac{{\rm i}}{2N_c} {\rm Tr}\big( \Omega^{A \dagger}_{\mu}(n; \{  a_{n,\mu,\nu,s} \} ) -\Omega^{A}_{\mu}(n; \{ a_{n,\mu,\nu,s} \} ) \big), 
\end{align}
and
\begin{align}
       \Omega^{A}_{\mu}(n; \{a_{n,\mu,\nu,s} \} ) &= C^{A}_{\mu}(n; \{ a_{n,\mu,\nu,s}  \} ) U_{\mu}^{(V)\dagger}(n),\notag\\
 C^{A}_{\mu}(n; \{ a_{n,\mu,\nu,s} \} ) &=\sum_{\nu \ne \mu}   \sum_{s=1}^{R} a_{n,\mu,\nu,s}  (U_{\nu}^{(V)}(n) U_{\mu}^{(V)}(n+\hat{\nu})U_{\nu}^{(V) \dagger}(n + \hat{\mu}) \\&+U^{(V)\dagger}_{\nu}(n-\hat{\nu})U^{(V)}_{\mu}(n-\hat{\nu}) U_{\nu}^{(V)}(x-\hat{\nu} + \hat{\mu}).
\end{align}
Here $R$ is the same variable in Figure 1.

Training is done with backprop as the gauge covariant neural network, which is an extension of \cite{Morningstar:2003gk,Nagai:2021bhh}.

\subsection{Self-learning Hybrid Monte Carlo}
To examine the expressibility of the Transformer, we perform self-learning Hybrid Monte Carlo (SLHMC) \cite{Nagai_2020,Nagai:2021bhh}, which incorporates an approximated model. In SLHMC, two different actions are involved: one is the exact action governing the target system, and the other is an approximate action used to guide the evolution. The acceptance rate in SLHMC is given by
$\min\left(
1,
\exp[-(H' - H)]
\right)$ where the primed quantities indicate the updated configuration. SLHMC employs the approximate action in the molecular dynamics evolution, and, because the molecular dynamics trajectory is invertible, one can still perform a conventional accept-reject step based on the actual HMC Hamiltonian difference. As a result, we can get exact expectation values.

\section{Lattice setup}
In order to test the expressivity of CASK, we carry out simulations in ${\rm SU}(2)$ lattice gauge theory with dynamical fermions with SLHMC. As a proof-of-principle study, we use a $4^4$ lattice at gauge coupling $\beta = 2.7$ and employ naive staggered fermions with mass ${m} = 0.3$. CASK is utilized to represent the effective action in SLHMC. Both the exact gauge action and the fermion action match those of the target system. However, during the molecular dynamics evolution, we replace the fermion mass ${m} = 0.3$ by a different mass ${m}^{\rm eff} = 0.4$ in the effective action and use CASK links in place of the thin links. CASK here relies solely on the plaquette kernel for smearing, but introduces three neighboring rectangular Wilson loops in the attention block to capture extended correlations. The intention is that CASK links absorb the difference arising from the modified Dirac operator.

All simulations are implemented using \texttt{Gaugefields.jl} and \texttt{LatticeDiracOperators.jl} in JuliaQCD \cite{Nagai:2024yaf}, which are written in the Julia language \cite{bezanson2012juliafastdynamiclanguage}. The network parameters in CASK are trained via the Adam optimizer \cite{kingma2017adammethodstochasticoptimization}.

\section{Results}
\begin{figure}[t]
\begin{center}
\includegraphics[width=0.32\textwidth]{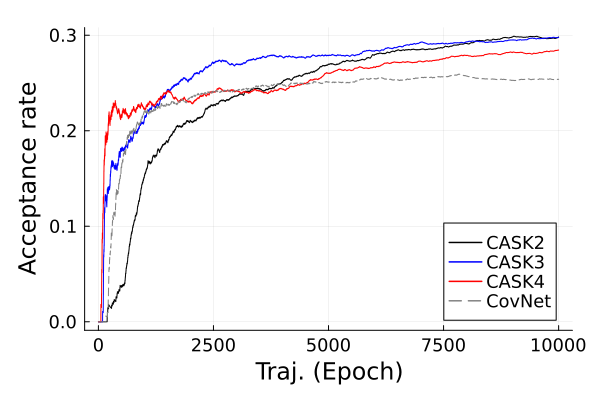}
\includegraphics[width=0.32\textwidth]{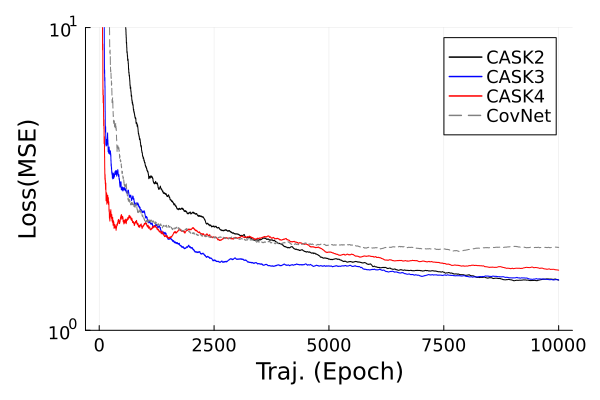}
\includegraphics[width=0.32\textwidth]{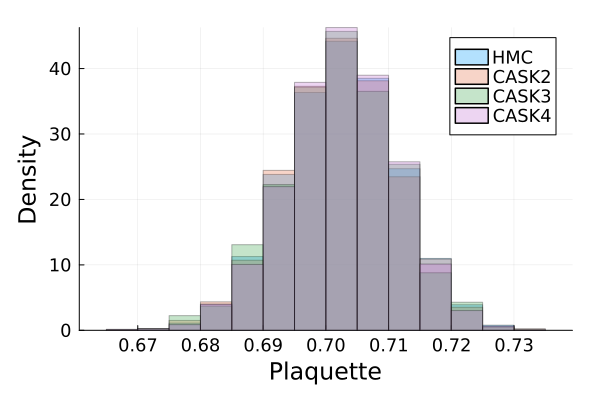}
\end{center}
\caption{
Comparison of algorithms.
({\it Left})
History of acceptance rate.
({\it Middle})
History of estimated loss function.
({\it Right})
Histogram of plaquette.
\label{fig:results}}
\end{figure}

We present our numerical findings in Fig.~\ref{fig:results}. In our setup, the SLHMC algorithm employs a Metropolis Hamiltonian $S = S_g + S_f[U, m]$ for the accept-reject step, while the molecular dynamics evolution uses an effective action $S = S_g + S_f[U^{\mathrm{eff}}, m^{\mathrm{eff}}]$. Here, $U^{\mathrm{eff}}$ is generated by a gauge covariant Transformer, and $m^{\mathrm{eff}}$ differs from the target fermion mass. Each Monte Carlo step corresponds to one epoch of training, during which the network parameters in the gauge covariant Transformer are updated. Without any training, the acceptance rate in this self-learning scheme is nearly zero, so any nonzero acceptance demonstrates that the network has sufficient expressive power to approximate the difference between $S_f[U, m]$ and $S_f[U^{\mathrm{eff}}, m^{\mathrm{eff}}]$.

The left panel of Fig.~\ref{fig:results} shows 
the acceptance rate and the middle panel is the estimated loss function as a function of the epoch by a formula in \cite{Nagai:2023fxt}. 
Different colors represent distinct setups.
CovNet indicates a result of the gauge covariant network and  CASK$n$ is CASK with $n$ attention blocks.
Over successive epochs, all networks learn to decrease the loss function. The gauge covariant neural network eventually saturates and ceases to improve, whereas the gauge covariant Transformer continues to learn at later epochs, achieving lower loss. This behavior highlights the Transformer’s enhanced capability to capture non-local correlations and model the effective action more flexibly. 

The right panel of Fig.~\ref{fig:results} illustrates that, even while the Transformer’s acceptance rate continues to grow, key observables remain consistent with expected physical behavior. This consistency confirms that the learned surrogate action does not distort essential physics, underscoring the practicality and reliability of gauge covariant Transformers in SLHMC.

\section{Summary}
In this work, we introduced the gauge covariant Transformer architecture CASK and demonstrated its utility in SLHMC simulations. By combining the essential features of gauge covariant neural networks with Transformer-based attention, CASK effectively incorporates both gauge symmetry and non-local correlations. In our numerical experiments, the surrogate links generated by CASK successfully absorbed the differences arising from the modified massive Dirac operator, resulting in an improved acceptance rate. 
The method consistently outperformed the gauge covariant neural networks (``adaptive stout'') developed in our previous study, illustrating how the attention-based design can enhance expressivity. These findings suggest that the gauge covariant Transformer approach is a promising route toward more efficient and flexible simulations in lattice QCD.
Future work will explore larger lattice volumes, extended loop structures in the attention matrix, and further optimization of the training process to fully leverage the potential of CASK.

\section*{Aknowledgement}
The work of authors was partially supported by JSPS KAKENHI Grant Numbers 20K14479,
22K03539, 22H05112, and 22H05111, and MEXT as “Program for Promoting Researches on the Supercomputer
Fugaku” (Simulation for basic science: approaching the new quantum era; Grant Number JPMXP1020230411, and
Search for physics beyond the standard model using large-scale lattice QCD simulation and development of AI
technology toward next-generation lattice QCD; Grant Number JPMXP1020230409).
Numerical computation in this work was partially carried out at the 
Yukawa Institute Computer Facility.

\bibliographystyle{utphys}
\bibliography{ref}

\providecommand{\href}[2]{#2}\begingroup\raggedright\begin{thebibliography}{10}

\bibitem{Adelmann:2022ozp}
A.~Adelmann {\em et~al.}, ``{New directions for surrogate models and differentiable programming for High Energy Physics detector simulation},'' in {\em {Snowmass 2021}}.
\newblock 3, 2022.
\newblock \href{http://arxiv.org/abs/2203.08806}{{\ttfamily arXiv:2203.08806 [hep-ph]}}.

\bibitem{Marcos_2017}
D.~Marcos, M.~Volpi, N.~Komodakis, and D.~Tuia, \href{http://dx.doi.org/10.1109/iccv.2017.540}{``Rotation equivariant vector field networks,''} in {\em 2017 IEEE International Conference on Computer Vision (ICCV)}.
\newblock IEEE, Oct., 2017.
\newblock \url{http://dx.doi.org/10.1109/ICCV.2017.540}.

\bibitem{Albergo:2019eim}
M.~S. Albergo, G.~Kanwar, and P.~E. Shanahan, ``{Flow-based generative models for Markov chain Monte Carlo in lattice field theory},'' \href{http://dx.doi.org/10.1103/PhysRevD.100.034515}{{\em Phys. Rev. D} {\bfseries 100} no.~3, (2019) 034515}, \href{http://arxiv.org/abs/1904.12072}{{\ttfamily arXiv:1904.12072 [hep-lat]}}.

\bibitem{Kanwar:2020xzo}
G.~Kanwar, M.~S. Albergo, D.~Boyda, K.~Cranmer, D.~C. Hackett, S.~Racani\`ere, D.~J. Rezende, and P.~E. Shanahan, ``{Equivariant flow-based sampling for lattice gauge theory},'' \href{http://dx.doi.org/10.1103/PhysRevLett.125.121601}{{\em Phys. Rev. Lett.} {\bfseries 125} no.~12, (2020) 121601}, \href{http://arxiv.org/abs/2003.06413}{{\ttfamily arXiv:2003.06413 [hep-lat]}}.

\bibitem{Boyda:2020hsi}
D.~Boyda, G.~Kanwar, S.~Racani\`ere, D.~J. Rezende, M.~S. Albergo, K.~Cranmer, D.~C. Hackett, and P.~E. Shanahan, ``{Sampling using $SU(N)$ gauge equivariant flows},'' \href{http://dx.doi.org/10.1103/PhysRevD.103.074504}{{\em Phys. Rev. D} {\bfseries 103} no.~7, (2021) 074504}, \href{http://arxiv.org/abs/2008.05456}{{\ttfamily arXiv:2008.05456 [hep-lat]}}.

\bibitem{Albergo:2021bna}
M.~S. Albergo, G.~Kanwar, S.~Racani\`ere, D.~J. Rezende, J.~M. Urban, D.~Boyda, K.~Cranmer, D.~C. Hackett, and P.~E. Shanahan, ``{Flow-based sampling for fermionic lattice field theories},'' \href{http://dx.doi.org/10.1103/PhysRevD.104.114507}{{\em Phys. Rev. D} {\bfseries 104} no.~11, (2021) 114507}, \href{http://arxiv.org/abs/2106.05934}{{\ttfamily arXiv:2106.05934 [hep-lat]}}.

\bibitem{Albergo:2022qfi}
M.~S. Albergo, D.~Boyda, K.~Cranmer, D.~C. Hackett, G.~Kanwar, S.~Racani\`ere, D.~J. Rezende, F.~Romero-L\'opez, P.~E. Shanahan, and J.~M. Urban, ``{Flow-based sampling in the lattice Schwinger model at criticality},'' \href{http://dx.doi.org/10.1103/PhysRevD.106.014514}{{\em Phys. Rev. D} {\bfseries 106} no.~1, (2022) 014514}, \href{http://arxiv.org/abs/2202.11712}{{\ttfamily arXiv:2202.11712 [hep-lat]}}.

\bibitem{Abbott:2022zhs}
R.~Abbott {\em et~al.}, ``{Gauge-equivariant flow models for sampling in lattice field theories with pseudofermions},'' \href{http://dx.doi.org/10.1103/PhysRevD.106.074506}{{\em Phys. Rev. D} {\bfseries 106} no.~7, (2022) 074506}, \href{http://arxiv.org/abs/2207.08945}{{\ttfamily arXiv:2207.08945 [hep-lat]}}.

\bibitem{Abbott:2024kfc}
R.~Abbott, A.~Botev, D.~Boyda, D.~C. Hackett, G.~Kanwar, S.~Racani\`ere, D.~J. Rezende, F.~Romero-L\'opez, P.~E. Shanahan, and J.~M. Urban, ``{Applications of flow models to the generation of correlated lattice QCD ensembles},'' \href{http://dx.doi.org/10.1103/PhysRevD.109.094514}{{\em Phys. Rev. D} {\bfseries 109} no.~9, (2024) 094514}, \href{http://arxiv.org/abs/2401.10874}{{\ttfamily arXiv:2401.10874 [hep-lat]}}.

\bibitem{deHaan:2021erb}
P.~de~Haan, C.~Rainone, M.~C.~N. Cheng, and R.~Bondesan, ``{Scaling Up Machine Learning For Quantum Field Theory with Equivariant Continuous Flows},'' \href{http://arxiv.org/abs/2110.02673}{{\ttfamily arXiv:2110.02673 [cs.LG]}}.

\bibitem{Gerdes:2022eve}
M.~Gerdes, P.~de~Haan, C.~Rainone, R.~Bondesan, and M.~C.~N. Cheng, ``{Learning lattice quantum field theories with equivariant continuous flows},'' \href{http://dx.doi.org/10.21468/SciPostPhys.15.6.238}{{\em SciPost Phys.} {\bfseries 15} no.~6, (2023) 238}, \href{http://arxiv.org/abs/2207.00283}{{\ttfamily arXiv:2207.00283 [hep-lat]}}.

\bibitem{Gerdes:2024rjk}
M.~Gerdes, P.~de~Haan, R.~Bondesan, and M.~C.~N. Cheng, ``{Continuous normalizing flows for lattice gauge theories},'' \href{http://arxiv.org/abs/2410.13161}{{\ttfamily arXiv:2410.13161 [hep-lat]}}.

\bibitem{Bulgarelli:2024yrz}
A.~Bulgarelli, E.~Cellini, K.~Jansen, S.~K\"uhn, A.~Nada, S.~Nakajima, K.~A. Nicoli, and M.~Panero, ``{Flow-based Sampling for Entanglement Entropy and the Machine Learning of Defects},'' \href{http://arxiv.org/abs/2410.14466}{{\ttfamily arXiv:2410.14466 [quant-ph]}}.

\bibitem{Bulgarelli:2024cqc}
A.~Bulgarelli, E.~Cellini, and A.~Nada, ``{Sampling SU(3) pure gauge theory with Stochastic Normalizing Flows},'' in {\em {41st International Symposium on Lattice Field Theory}}.
\newblock 9, 2024.
\newblock \href{http://arxiv.org/abs/2409.18861}{{\ttfamily arXiv:2409.18861 [hep-lat]}}.

\bibitem{Caselle:2024ent}
M.~Caselle, E.~Cellini, and A.~Nada, ``{Numerical determination of the width and shape of the effective string using Stochastic Normalizing Flows},'' \href{http://arxiv.org/abs/2409.15937}{{\ttfamily arXiv:2409.15937 [hep-lat]}}.

\bibitem{Holland:2024muu}
K.~Holland, A.~Ipp, D.~I. M\"uller, and U.~Wenger, ``{Machine learning a fixed point action for SU(3) gauge theory with a gauge equivariant convolutional neural network},'' \href{http://dx.doi.org/10.1103/PhysRevD.110.074502}{{\em Phys. Rev. D} {\bfseries 110} no.~7, (2024) 074502}, \href{http://arxiv.org/abs/2401.06481}{{\ttfamily arXiv:2401.06481 [hep-lat]}}.

\bibitem{Lehner:2023bba}
C.~Lehner and T.~Wettig, ``{Gauge-equivariant neural networks as preconditioners in lattice QCD},'' \href{http://dx.doi.org/10.1103/PhysRevD.108.034503}{{\em Phys. Rev. D} {\bfseries 108} no.~3, (2023) 034503}, \href{http://arxiv.org/abs/2302.05419}{{\ttfamily arXiv:2302.05419 [hep-lat]}}.

\bibitem{Lehner:2023prf}
C.~Lehner and T.~Wettig, ``{Gauge-equivariant pooling layers for preconditioners in lattice QCD},'' \href{http://dx.doi.org/10.1103/PhysRevD.110.034517}{{\em Phys. Rev. D} {\bfseries 110} no.~3, (2024) 034517}, \href{http://arxiv.org/abs/2304.10438}{{\ttfamily arXiv:2304.10438 [hep-lat]}}.

\bibitem{Cali:2022qbd}
S.~Cal\`\i{}, D.~C. Hackett, Y.~Lin, P.~E. Shanahan, and B.~Xiao, ``{Neural-network preconditioners for solving the Dirac equation in lattice gauge theory},'' \href{http://dx.doi.org/10.1103/PhysRevD.107.034508}{{\em Phys. Rev. D} {\bfseries 107} no.~3, (2023) 034508}, \href{http://arxiv.org/abs/2208.02728}{{\ttfamily arXiv:2208.02728 [hep-lat]}}.

\bibitem{cuomo2022scientificmachinelearningphysicsinformed}
S.~Cuomo, V.~S. di~Cola, F.~Giampaolo, G.~Rozza, M.~Raissi, and F.~Piccialli, ``Scientific machine learning through physics-informed neural networks: Where we are and what's next,'' 2022.
\newblock \url{https://arxiv.org/abs/2201.05624}.

\bibitem{ruder2017overviewgradientdescentoptimization}
S.~Ruder, ``An overview of gradient descent optimization algorithms,'' 2017.
\newblock \url{https://arxiv.org/abs/1609.04747}.

\bibitem{blondel2024elementsdifferentiableprogramming}
M.~Blondel and V.~Roulet, ``The elements of differentiable programming,'' 2024.
\newblock \url{https://arxiv.org/abs/2403.14606}.

\bibitem{moses2020instead}
W.~Moses and V.~Churavy, ``Instead of rewriting foreign code for machine learning, automatically synthesize fast gradients,'' {\em Advances in neural information processing systems} {\bfseries 33} (2020) 12472--12485.

\bibitem{vaswani2023attentionneed}
A.~Vaswani, N.~Shazeer, N.~Parmar, J.~Uszkoreit, L.~Jones, A.~N. Gomez, L.~Kaiser, and I.~Polosukhin, ``Attention is all you need,'' 2023.
\newblock \url{https://arxiv.org/abs/1706.03762}.

\bibitem{lin2021surveytransformers}
T.~Lin, Y.~Wang, X.~Liu, and X.~Qiu, ``A survey of transformers,'' 2021.
\newblock \url{https://arxiv.org/abs/2106.04554}.

\bibitem{brehmer2024doesequivariancematterscale}
J.~Brehmer, S.~Behrends, P.~de~Haan, and T.~Cohen, ``Does equivariance matter at scale?,'' 2024.
\newblock \url{https://arxiv.org/abs/2410.23179}.

\bibitem{Nagai:2021bhh}
Y.~Nagai and A.~Tomiya, ``{Gauge covariant neural network for quarks and gluons},'' \href{http://arxiv.org/abs/2103.11965}{{\ttfamily arXiv:2103.11965 [hep-lat]}}.

\bibitem{Nagai:2023fxt}
Y.~Nagai and A.~Tomiya, ``{Self-learning Monte Carlo with equivariant Transformer},'' \href{http://dx.doi.org/10.7566/JPSJ.93.114007}{{\em J. Phys. Soc. Jap.} {\bfseries 93} (2024) 114007}, \href{http://arxiv.org/abs/2306.11527}{{\ttfamily arXiv:2306.11527 [cond-mat.str-el]}}.

\bibitem{Tomiya:2023jdy}
A.~Tomiya and Y.~Nagai, ``{Equivariant transformer is all you need},'' \href{http://dx.doi.org/10.22323/1.453.0001}{{\em PoS} {\bfseries LATTICE2023} (2024) 001}, \href{http://arxiv.org/abs/2310.13222}{{\ttfamily arXiv:2310.13222 [hep-lat]}}.

\bibitem{belilovsky2019greedylayerwiselearningscale}
E.~Belilovsky, M.~Eickenberg, and E.~Oyallon, ``Greedy layerwise learning can scale to imagenet,'' 2019.
\newblock \url{https://arxiv.org/abs/1812.11446}.

\bibitem{child2019generatinglongsequencessparse}
R.~Child, S.~Gray, A.~Radford, and I.~Sutskever, ``Generating long sequences with sparse transformers,'' 2019.
\newblock \url{https://arxiv.org/abs/1904.10509}.

\bibitem{Morningstar:2003gk}
C.~Morningstar and M.~J. Peardon, ``{Analytic smearing of SU(3) link variables in lattice QCD},'' \href{http://dx.doi.org/10.1103/PhysRevD.69.054501}{{\em Phys. Rev. D} {\bfseries 69} (2004) 054501}, \href{http://arxiv.org/abs/hep-lat/0311018}{{\ttfamily arXiv:hep-lat/0311018}}.

\bibitem{Nagai_2020}
Y.~Nagai, M.~Okumura, K.~Kobayashi, and M.~Shiga, ``Self-learning hybrid monte carlo: A first-principles approach,'' \href{http://dx.doi.org/10.1103/physrevb.102.041124}{{\em Physical Review B} {\bfseries 102} no.~4, (July, 2020) }. \url{http://dx.doi.org/10.1103/PhysRevB.102.041124}.

\bibitem{Nagai:2024yaf}
Y.~Nagai and A.~Tomiya, ``{JuliaQCD: Portable lattice QCD package in Julia language},'' \href{http://arxiv.org/abs/2409.03030}{{\ttfamily arXiv:2409.03030 [hep-lat]}}.

\bibitem{bezanson2012juliafastdynamiclanguage}
J.~Bezanson, S.~Karpinski, V.~B. Shah, and A.~Edelman, ``Julia: A fast dynamic language for technical computing,'' 2012.
\newblock \url{https://arxiv.org/abs/1209.5145}.

\bibitem{kingma2017adammethodstochasticoptimization}
D.~P. Kingma and J.~Ba, ``Adam: A method for stochastic optimization,'' 2017.
\newblock \url{https://arxiv.org/abs/1412.6980}.

\end{thebibliography}\endgroup

\end{document}